# MESHLESS METHOD FOR SOLVING RADIATIVE TRANSFER PROBLEMS IN COMPLEX TWO-DIMENSIONAL AND THREE-DIMENSIONAL GEOMETRIES

**Cheng-An Wang, Hamou Sadat, Vital Ledez and Denis Lemonnier**

Laboratoire d'Etudes Thermiques, UMR 6608, University of Poitiers,
40 Avenue du Recteur Pineau, Poitiers cedex, France
E-mail : hamou.sadat@univ-poitiers.fr

**Abstract:**

A meshless method is presented to solve the radiative transfer equation in the even parity formulation of the discrete ordinates method in complex 2D and 3D geometries. Prediction results of radiative heat transfer problems obtained by the proposed method are compared with reference in order to assess the correctness of the present method.

## 1. Introduction

In the last decade, a large family of meshless methods have been developed for solving partial differential equations. Comprehensive literature surveys and textbooks may be consulted for reference [1-3]. In this paper, we focus on the moving least squares collocation meshless method whose origin can be found in different independent works. The notion of moving least squares approximation has been introduced to improve the least squares approximation in the early seventies for generating surfaces [4]. In order to improve the finite element interpolation, diffuse approximation which is closely related to MLS has been presented [5] and used in a Galerkin-like approach in what was called the diffuse element method [6] and has been further improved in the development of the Element Free Galerkin method [7].

Rather than introducing the MLS approximation in a Galerkin procedure, one can use it in a collocation approach to approximate the derivatives of the unknown scalar field in a cloud of points and solve partial differential equations. One of the first attempts to solve heat transfer and fluid flow problems by using a diffuse approximation or a moving least squares collocation method was described in [8] where fluid flow and natural convection problems have been solved in the vorticity-streamfunction formulation. The method has been found efficient and comparable in accuracy to the well established control volume based finite element method [9]. By using a projection algorithm for the pressure velocity coupling, fluid flow problems have been solved in the primitive variable formulation of the Navier Stokes equations [10].Unsteady two and three dimensional problems have been considered and the method was shown to be accurate both in space and time [11-12]. The method has also been compared to classical methods when solving a convection dominated phase change benchmark problem [13]. It has been found that it was accurate even for the high values of the considered Rayleigh numbers. Further it was one of the only two methods which predicted a multicellular flow in one test case.

There exist numerous engineering situations where heat transfer and fluid flow occur in a semi transparent medium. In these cases, radiative transfer must be solved together with other transfer processes in an iterative manner on the same mesh or set of nodes. Thus, any new numerical method must also be able to accurately calculate radiative transfer. The main goal of this article is therefore to show that the diffuse approximation or the moving least squares collocation method can tackle radiative transfer problems with a sufficient accuracy.
Due to absorption and diffusion phenomena in the participating medium, radiative intensity which is the fundamental quantity in this field obeys an integro-differential equation called the radiative transfer equation (RTE) whose main difficulty is its directional nature. If one excepts Monte Carlo and Ray Tracing methods which are time consuming and not easily coupled with, two classes of methods can be used, namely the spherical harmonics method also called moment method or simply $P_N$ method and the discrete ordinates method (DOM) or $S_N$ method. The $P_N$ method uses an approximation of the radiative intensity and converts the integro-differential RTE into an Helmholtz type equation which can be solved by standard numerical methods. In the Discrete ordinates method, the RTE is converted into a system of partial differential equations which can also be solved with standard numerical methods such as finite volume [14-16] or finite element [17-18] methods. Application of the even parity formulation of the RTE to the DOM has been implemented by several researchers [19-20]. The second-order differential form of the even parity formulation is attractive since it reduces the number of governing equations to half compared with the conventional first-order discrete ordinates equation. Also, the resulting equations are familiar second-order differential equations.

Some attempts to solve radiative transfer problems by using meshless methods have been proposed. Two dimensional geometries have been mainly considered [21-23]. In a previous paper [24] we have shown that the even parity formulation has better stability when using a collocation meshless method than the primitive variables formulation. In this work, we extend the approach used in [24] to two dimensional and three dimensional complex geometry problems. This paper is organized as follows. The discrete ordinates method in the primitive variables and the even parity formulations are first described. The collocation meshless method and the discretization of the equations are then presented. To evaluate the accuracy and computational efficiency of the present meshless method, various 2D and 3D test cases are finally considered. The results obtained are discussed and compared with other available solutions.

## 2. The discrete ordinates method

The DOM is based on the use of numerical quadratures to approximate the integrals that appear in the calculation of the incident radiation and partial heat fluxes. It uses a discretization of the angular space by a finite number of directions along which the RTE is solved. In the primitive variables formulation, considering a discrete ordinate $i$ with coordinate $s$, for an absorbing, emitting and scattering medium, the RTE is written as:

$$\frac{dI(\Omega_i)}{ds} = -\beta I(\Omega_i) + \kappa I_b + \frac{\sigma}{4\pi} \sum_{j=1}^{J} I(\Omega'_j) \Phi(\Omega'_j, \Omega_i) W(\Omega'_j) \tag{1}$$

where $j = 1...J$ are the discrete ordinates directions, and $\kappa$, $\sigma$ and $\beta$ are the absorption, the scattering and the extinction coefficients, respectively, $I_b$ is the blackbody intensity of the medium, $\Phi(\Omega'_j, \Omega_i)$ is the scattering phase function of intensity entering from $\Omega'_j$ scattered to $\Omega_i$ and $W(\Omega'_j)$ is the angular weight of ordinate $j$.

Boundary condition for diffuse walls are:

$$I(\Omega_i) = \varepsilon I_{bw} + \frac{1-\varepsilon}{\pi} \sum_{\hat{n}\cdot\Omega'_j<0} I(\Omega'_j)|\hat{n}\cdot\Omega'_j|W(\Omega'_j) \tag{2}$$

where $\varepsilon$ is the wall emissivity, $I_{bw}$ is the blackbody intensity of the wall, $\hat{n}$ is the unit inward normal vector at the boundary location and $\Omega'_j$ is the unit direction vector of $j$ th ordinate.

Basically, the discrete ordinates (or $S_N$) method proceeds as follows:

1. The RTE is solved for a discrete set of directions $\Omega_i$, $i=1...J$. Then, for each direction $i$, $I(\Omega_i)$ is known in the whole calculation domain
2. Weighting factors $W(\Omega_i)$ are associated with each direction $\Omega_i$ so that integration over directions may be approximated by the general quadrature formula:

$$\int_{4\pi} I(\bar{\Omega})d\Omega = \sum_{i=1}^{J} I(\bar{\Omega})W(\Omega_i) \tag{3}$$

## 3. Even-parity formulation of the discrete ordinates method

The second-order formulation begins with taking two intensities in opposite directions. The intensities to the positive and the negative directions are denoted as $I^+(\Omega)$ and $I^-(\Omega)$ and are governed by the following RTEs:

$$\frac{dI^+(\Omega)}{ds} = -\beta I^+(\Omega) + \kappa I_b + \frac{\sigma}{4\pi}\int_{4\pi} I^+(\Omega')\Phi(\Omega',\Omega)d\Omega' \tag{4}$$

$$\frac{dI^-(\Omega)}{ds} = \beta I^-(\Omega) - \kappa I_b - \frac{\sigma}{4\pi}\int_{4\pi} I^-(\Omega')\Phi(\Omega',-\Omega)d\Omega' \tag{5}$$

Introducing the following new variables defined as:

$$F(\Omega) = I^+(\Omega) + I^-(\Omega) \tag{6}$$
$$G(\Omega) = I^+(\Omega) - I^-(\Omega) \tag{7}$$

and assuming that the scattering-phase function satisfies the following general condition:

$$\Phi(\Omega',\Omega) = \Phi(-\Omega',-\Omega) \tag{8}$$

It can be shown that one can eliminate one of the variables and write for example for the variable $F_i$ (calculated on each line of sight):

$$\frac{1}{\beta}\frac{d^2 F_i}{ds^2} - \beta F_i + \kappa I_b + \frac{\sigma}{4\pi}\sum_{j=1}^{J/2}(A_{ij}F_j + B_{ij}G_j) = 0 \qquad (9)$$

Where :

$$A_{ij} = [\Phi(\Omega'_j, \Omega_i) + \Phi(\Omega'_j, -\Omega_i)]W(\Omega'_j) \qquad (10)$$

and

$$B_{ij} = [\Phi(\Omega'_j, \Omega_i) - \Phi(\Omega'_j, -\Omega_i)]W(\Omega'_j) \qquad (11)$$

The corresponding boundary condition is:

$$F_i - sign(\hat{n}\cdot\Omega'_i)\frac{1}{\beta}\frac{dF_i}{ds} = \varepsilon I_{bw} + \frac{1-\varepsilon}{\pi}q' \qquad (12)$$

Where $q'$ is given by:

$$q' = \sum_{j=1, \hat{n}\cdot\Omega'_j<0}^{J}(F_i + G_i)|\hat{n}\cdot\Omega'_j|W(\Omega'_j) \qquad (13)$$

Function $G$ can be deduced by using the equation :

$$\frac{\partial F_i(P,\vec{\Omega})}{\partial s_m} + \beta G_i(P,\vec{\Omega}) = 0 \qquad (14)$$

The incident radiation, the partial fluxes and the boundary condition on a wall are then calculated by the usual numerical quadratures.

## 4. The moving least squares collocation meshless method

The partial differential equations are solved by a moving least squares based meshless method whose main characteristics are presented in this section.

Let $\Phi : R^n \rightarrow R$ be a scalar field whose values $\Phi_i$ are known at the point $X_i$ of a given set of $N$ nodes in the studied domain $D \subset R^n$. The diffuse approximation gives estimates of $\Phi$ and its derivatives up to the order $k$ at any point $X \in D$. The Taylor expansion of $\Phi$ at $X$ is estimated by a weighted least squares method which uses only the values of $\Phi$ at some points $X_i$ situated in the vicinity of $X$.

It can thus be written:

$$\Phi_i^{estimated} = \langle P(X_i - X) \rangle \langle \alpha(X) \rangle^T \tag{15}$$

where $\langle P(X_i - X) \rangle$ is a colon vector of polynomial basis functions and $\langle \alpha(X) \rangle^T$ a vector of coefficients which are determined by minimizing the quantity:

$$I(\alpha) = \sum_{i=1}^{N} \omega(X, X_i - X) \left[ \Phi_i - \langle P(X_i - X) \rangle \langle \alpha(X) \rangle^T \right]^2 \tag{16}$$

in which $\omega(X, X_i - X)$ is a weight-function of compact support, equal to unity at this point, decreasing when the distance to the node increases and zero outside a given domain of influence (a more precise description of $\omega(X, X_i - X)$ will be done next).

Minimization of Eq. (16) then gives:

$$A(X)\alpha(X) = B(X) \tag{17}$$

where:

$$A(X) = \sum_{i=1}^{N} \omega(X, X_i - X) P(X_i - X) P^T(X_i - X) \tag{18}$$

$$B(X) = \sum_{i=1}^{N} \omega(X, X_i - X) P(X_i - X) \Phi_i \tag{19}$$

By inverting system (17), one obtains the components of $\alpha$ which are the derivatives of $\Phi$ at $X$ in terms of the neighboring nodal values $\Phi_i$. In this work, the Taylor expansion is truncated at order 2. The polynomial vector used is

$$\langle P(X_i - X) \rangle = \langle 1, (x_i - x), (y_i - y), (x_i - x)^2, (x_i - x) \cdot (y_i - y), (y_i - y)^2 \rangle \tag{20}$$

and we have:

$$\langle \alpha_1, \alpha_2, \alpha_3, \alpha_4, \alpha_5, \alpha_6 \rangle^T = \langle \Phi, \frac{\partial \Phi}{\partial x}, \frac{\partial \Phi}{\partial y}, \frac{\partial^2 \Phi}{2\partial x^2}, \frac{\partial^2 \Phi}{\partial x \partial y}, \frac{\partial^2 \Phi}{2\partial y^2} \rangle^T \tag{21}$$

Finally, the following system is obtained:

$$\begin{Bmatrix} \varphi \\ \dfrac{\partial \varphi}{\partial x} \\ \dfrac{\partial \varphi}{\partial y} \\ \dfrac{\partial^2 \varphi}{2!\partial x^2} \\ \dfrac{\partial^2 \varphi}{\partial x \partial y} \\ \dfrac{\partial^2 \varphi}{2!\partial y^2} \end{Bmatrix}^* = [A(X)]^{-1} \cdot \left\{ \sum_{i=1}^{n'(X)} \omega(X, X_i - X) \langle P(X_i - X) \rangle^T \cdot \Phi_i \right\} \tag{22}$$

The square matrix $A(X)$ is not singular as long as the number $n'(X)$ of the connected nodes at a given point is at least equal to the size of $\langle P(X_i - X) \rangle$ and are not all situated in the same plane (in 3D) or line (in 2D).

In our studies, several weight-functions were tried and it was found that the following Gaussian window :

$$\omega(X, X_i - X) = \exp\left[ -3\ln(10) \cdot \left( \frac{|X_i - X|}{\sigma} \right)^2 \right]$$
$$\omega(X, X_i - X) = 0 \quad \text{if } (X_i - X)^2 > \sigma^2 \tag{23}$$

behaves rather well. The distance of influence $\sigma$ is updated at each point. In our work, the distance of influence is chosen at each point to include 9 neighbours (in 2D) and 27 neighbours (in 3D).

The previous approximation is then used in a point collocation method to solve partial derivatives equations. At each point of the discretization, the derivatives appearing in the equation to be solved are replaced by their diffuse approximation thus leading to an algebraic system that is solved after the introduction of the boundary conditions.

The Dirichlet type boundary conditions are introduced in the same way as in the finite element method. The Neumann boundary conditions on the other hand are replaced by their diffuse approximation and then introduced in the algebraic system. The final algebraic system is solved by using the BICGSTAB iterative method.

## 5. Discretization schemes

In the following, we set $\langle p(M_j, M) \rangle = \langle p_j \rangle$ and we define $\langle a_i \rangle$ as the $i$ th line of the inverse matrix $[A^M]^{-1}$.

For 2D problems, the governing equation 9 can be written:

$$\frac{1}{\beta}[(\mu^m)^2 \frac{\partial^2 F_i}{\partial x^2} + (\eta^m)^2 \frac{\partial^2 F_i}{\partial y^2} + 2\mu^m \eta^m \frac{\partial^2 F_i}{\partial x \partial y}] - \beta F_i + \kappa I_b + \frac{\sigma}{4\pi} \sum_{j=1}^{J/2} (A_{ij} F_j + B_{ij} G_j) = 0 \quad (24)$$

Where $\mu^m$ and $\eta^m$ are the cosine directors of the line of sight $i$.

By using the explicit relations of the derivates given by Eq.(22), at each point of the calculation grid, one obtains the following algebraic system:

$$[MAT] \cdot [F_i] = [S] \quad (25)$$

Where we have:

$$MAT(k,j) = \omega(M_j, M)[2!\frac{1}{\beta}((\mu^m)^2\langle a_4\rangle + (\eta^m)^2\langle a_6\rangle + 2\mu^m\eta^m\langle a_5\rangle) - \beta\langle a_1\rangle] \cdot \langle p_j\rangle^T \quad (26)$$

and

$$S(k) = -\left(\kappa I_b + \frac{\sigma}{4\pi}\sum_{j=1}^{J/2}(A_{ij}F_j + B_{ij}G_j)\right) \quad (27)$$

On the other hand, Boundary equation Eq.(12) is written as:

$$F_i - sign(\hat{n}\cdot\Omega'_i)\frac{1}{\beta}[\mu^m\frac{\partial F_i}{\partial x} + \eta^m\frac{\partial F_i}{\partial y}] = \varepsilon I_{bw} + \frac{1-\varepsilon}{\pi}q' \quad (28)$$

For the points in the boundary, we therefore write:

$$MAT(k,j) = \omega(M_j, M)[\langle a_1\rangle - sign(\hat{n}\cdot\Omega'_i)\frac{1}{\beta}(\mu^m\langle a_2\rangle + \eta^m\langle a_3\rangle)] \cdot \langle p_j\rangle^T \quad (29)$$

and

$$S(k) = \varepsilon I_{bw} + \frac{1-\varepsilon}{\pi}q' \quad (30)$$

For 3D problems, the governing equation (Eq.9) is transformed as:

$$\frac{1}{\beta}[(\mu^m)^2\frac{\partial^2 F_i}{\partial x^2} + (\eta^m)^2\frac{\partial^2 F_i}{\partial y^2} + (\xi^m)^2\frac{\partial^2 F_i}{\partial z^2} + 2\mu^m\eta^m\frac{\partial^2 F_i}{\partial x\partial y} + 2\mu^m\xi^m\frac{\partial^2 F_i}{\partial x\partial z}$$
$$+ 2\eta^m\xi^m\frac{\partial^2 F_i}{\partial y\partial z}] - \beta F_i + \kappa I_b + \frac{\sigma}{4\pi}\sum_{j=1}^{J/2}(A_{ij}F_j + B_{ij}G_j) = 0 \quad (31)$$

Where $\mu^m, \eta^m, \xi^m$ are now the three components of the direction vector.

In this case, the coefficients of the Matrix and of the second member vector are written as:

$$MAT(k,j) = \omega(M_j, M)[2!\frac{1}{\beta}((\mu^m)^2\langle a_4\rangle + (\eta^m)^2\langle a_6\rangle + (\xi^m)^2\langle a_{10}\rangle +$$
$$2\mu^m\eta^m\langle a_5\rangle + 2\mu^m\xi^m\langle a_8\rangle + 2\eta^m\xi^m\langle a_9\rangle) - \beta\langle a_1\rangle] \cdot \langle p_j\rangle^T \quad (32)$$

$$S(k) = -\left(\kappa I_b + \frac{\sigma}{4\pi}\sum_{j=1}^{J/2}(A_{ij}F_j + B_{ij}G_j)\right) \quad (33)$$

Once again, Boundary equation Eq.(12) is written as:

$$F_i - sign(\hat{n}\cdot\Omega'_i)\frac{1}{\beta}[\mu^m\frac{\partial F_i}{\partial x} + \eta^m\frac{\partial F_i}{\partial y} + \zeta^m\frac{\partial F_i}{\partial z}] = \varepsilon I_{bw} + \frac{1-\varepsilon}{\pi}q' \quad (34)$$

For the points lying in the boundary, the following coefficients are obtained:

$$MAT(k,j) = \omega(M_j, M)[\langle a_1\rangle - sign(\hat{n}\cdot\Omega'_i)\frac{1}{\beta}(\mu^m\langle a_2\rangle + \eta^m\langle a_3\rangle + \xi^m\langle a_7\rangle)]\cdot\langle p_j\rangle^T \qquad (35)$$

$$S(k) = \varepsilon I_{bw} + \frac{1-\varepsilon}{\pi}q' \qquad (36)$$

## 6. Two-dimensional results

This section is devoted to the results obtained on some two-dimensional problems.

## 6. 1 Two-dimensional quadrilateral enclosure

Let us first consider here the example studied by Chai et al [25] in the geometry of Figure 1. This problem consists of an absorbing and emitting medium maintained at an emissive power of unity. The medium is enclosed by cold and black walls. The selected medium absorption coefficient $\kappa$ varies from 0.1 to 1.0 to 10 m$^{-1}$. This benchmark problem has been studied before and an exact solution for radiative heat flux along the bottom wall has been provided [25].

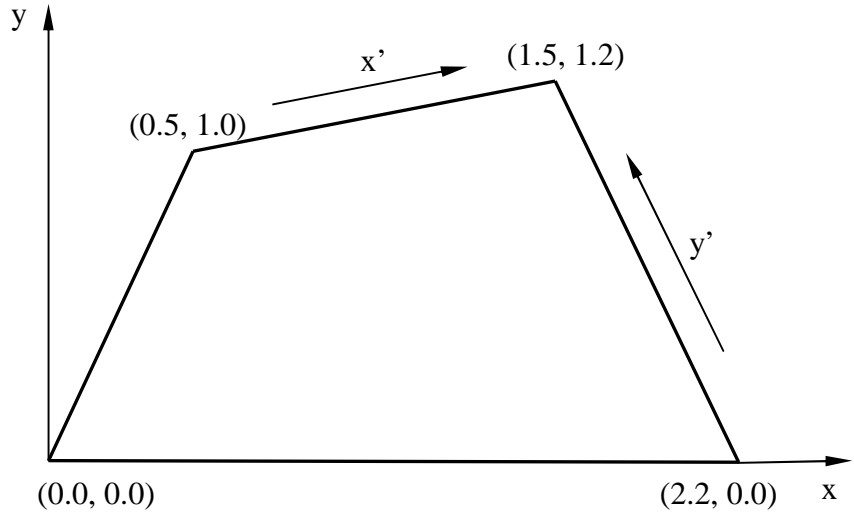

Figure 1 : Quadrilateral enclosure

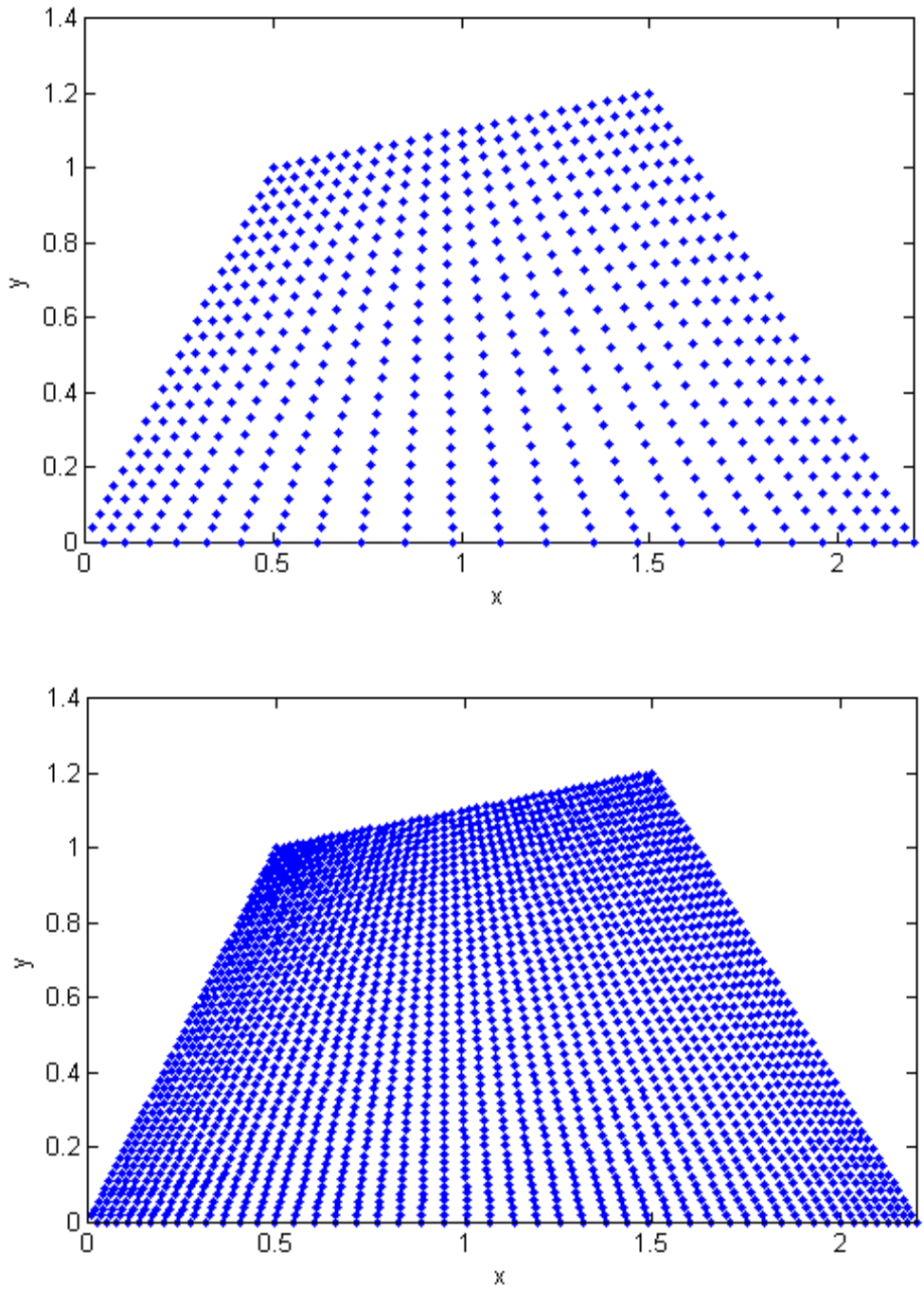

Figure 2 : example of nodes clouds:25*25 and 50*50 grids

Figure 2 shows two examples of grid points used for the calculations, namely 25*25 and 50*50. The radiative heat flux distributions on the bottom wall obtained with a 25*25 grid (for $\kappa$ =0.1 and 1 m$^{-1}$) and with a 50*50 grid (for $\kappa$ =10 m$^{-1}$) are presented on Figure 3 together with the exact solution. It can be seen that the results are correct.

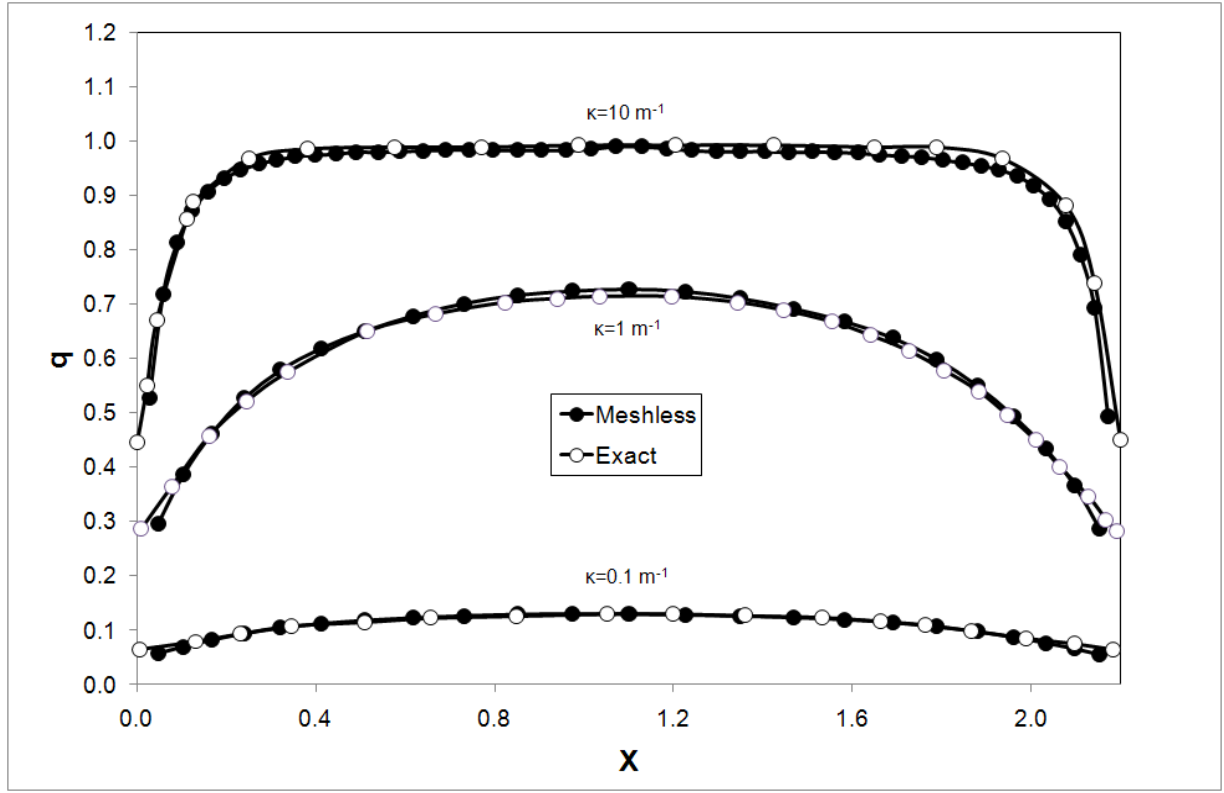

Figure 3: Heat flux distribution along the bottom wall

## 6.2 Quadrilateral enclosure: purely scattering case

Another problem has been considered in the same quadrilateral geometry. The lower wall of the quadrilateral is hot (T=300K), while the other walls are kept cold (T=0K). All the walls are black and the medium is cold and purely scattering. Our results obtained with S4 and S8 quadratures are compared to the Monte Carlo results which are the reference results [26]. We show on Figure 4 and 5, the radiative heat flux on the top and on the right walls obtained with the same spatial 25*25 grid. It is seen that the S8 quadrature leads to more accurate results than the S4 quadrature as expected.

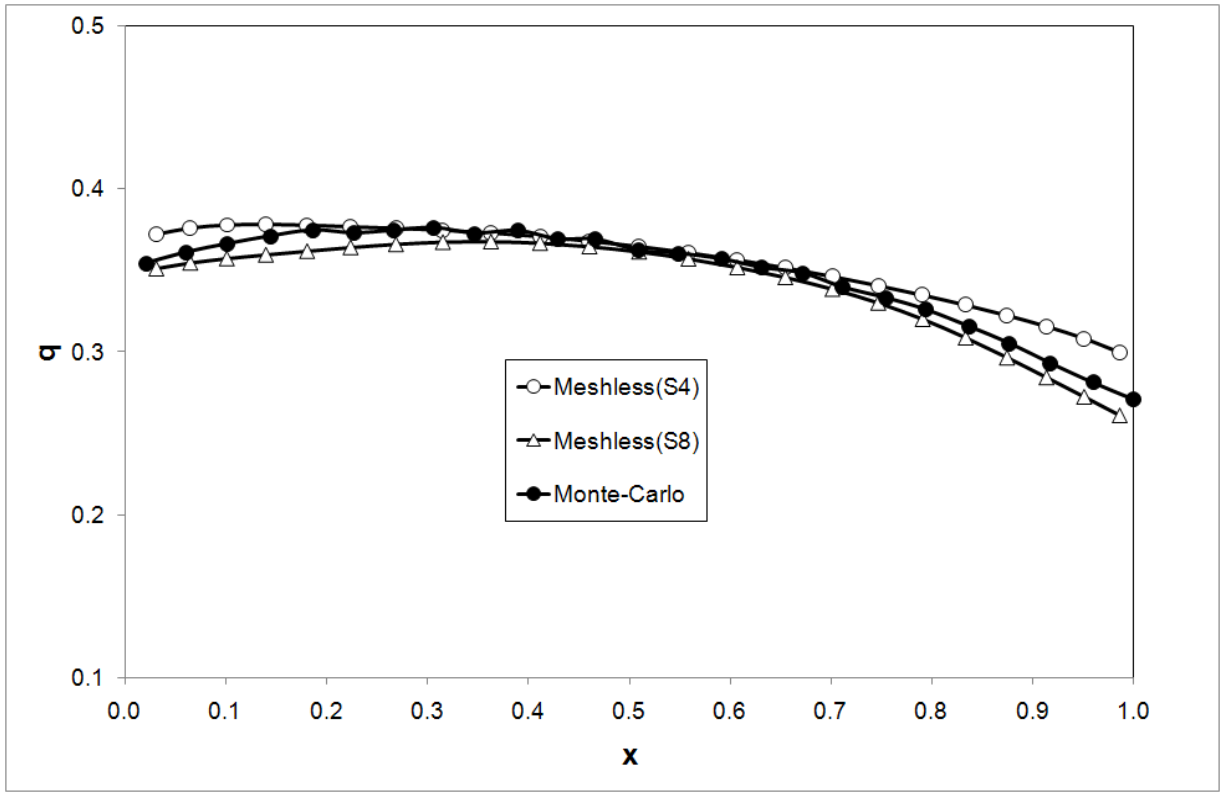

Figure 4: Flux distribution at the top boundary

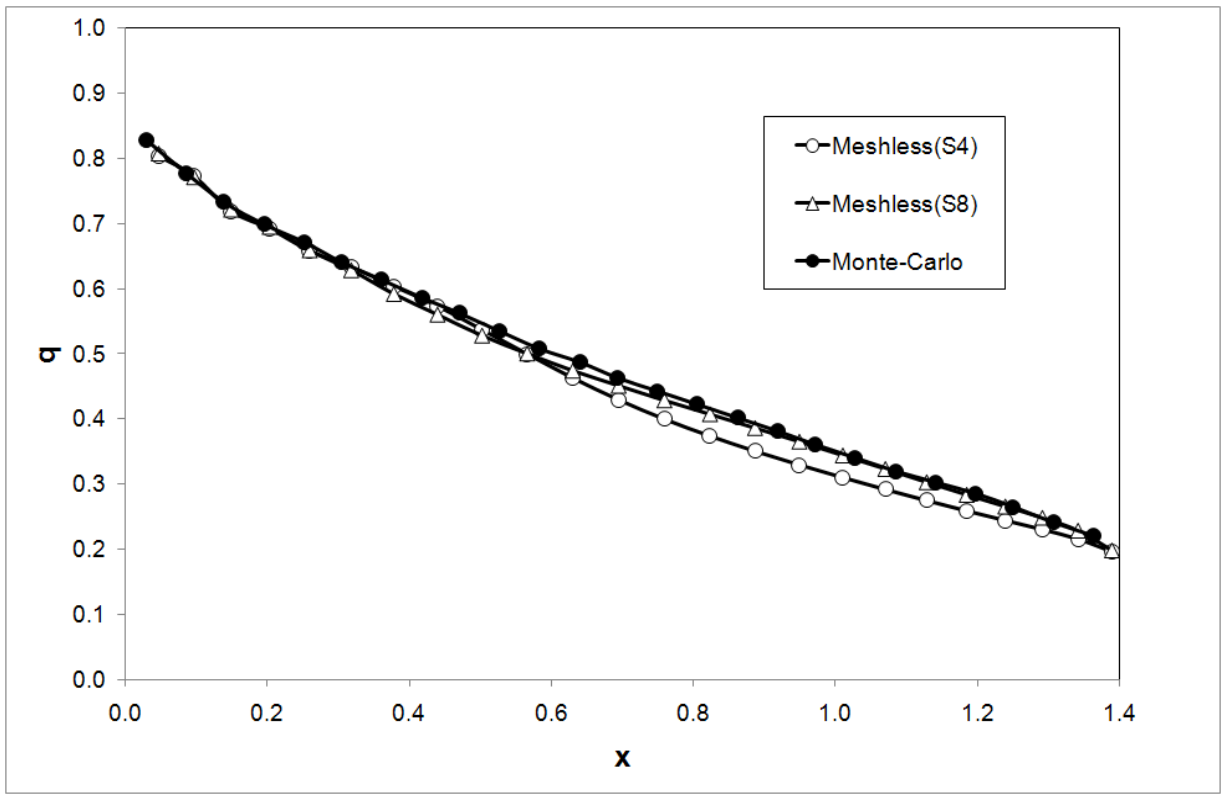

Figure 5: Flux distribution at the right boundary

Finally it is worth noting that (with a convergence test on the relative error equal than $10^{-5}$), the number of iterations necessary for convergence is equal to 17 for both quadratures while the CPU time with the S8 quadrature is 3.28 times greater than that with the S4 quadrature.

## 7. Three-dimensional results

We now turn to the presentation of three dimensional results obtained in an hexaedral, an L shaped and an elliptic enclosures respectively.

## 7.1 Hexahedral enclosure

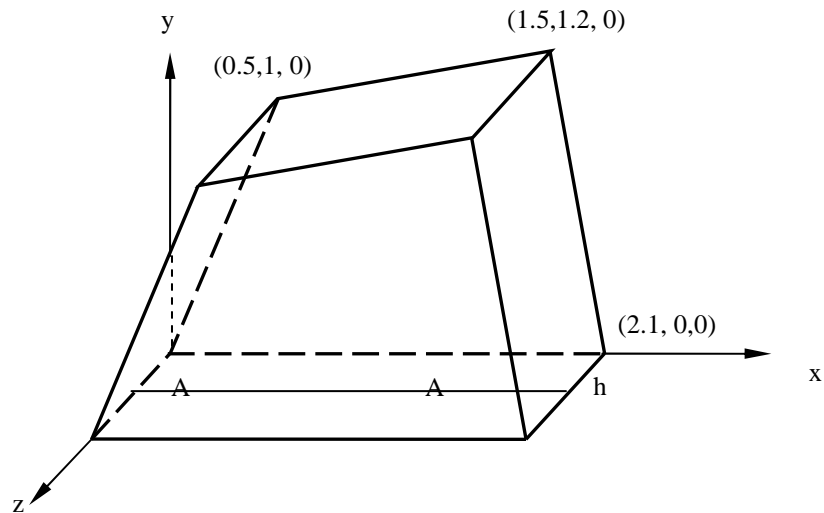

Figure 6: Hexahedral enclosure

The first problem deals with heat transfer in an absorbing and emitting semi-transparent medium maintained enclosed in the hexahedral enclosure shown on Figure 6(with $h = 1$m). The walls of the enclosure are assumed to be black and held at a constant zero temperature while the medium is at Tm=100K. Three different absorption coefficients are considered ($\kappa$=0.1, 1.0 and 10.0 m$^{-1}$). Our results are compared with the exact solution given in [27]. We show on Figure 7 the wall heat flux along the line (z=h/2) for the three values of the absorption coefficient and by using an $S_6$ approximation. The grids used in the calculations are composed by 567 nodes for $\kappa=0.1$, 2575 nodes for $\kappa=1$ and 92500 nodes for $\kappa=10$ m$^{-1}$. The maximum relative error for $\kappa=10$ m$^{-1}$ is lower than 2%.

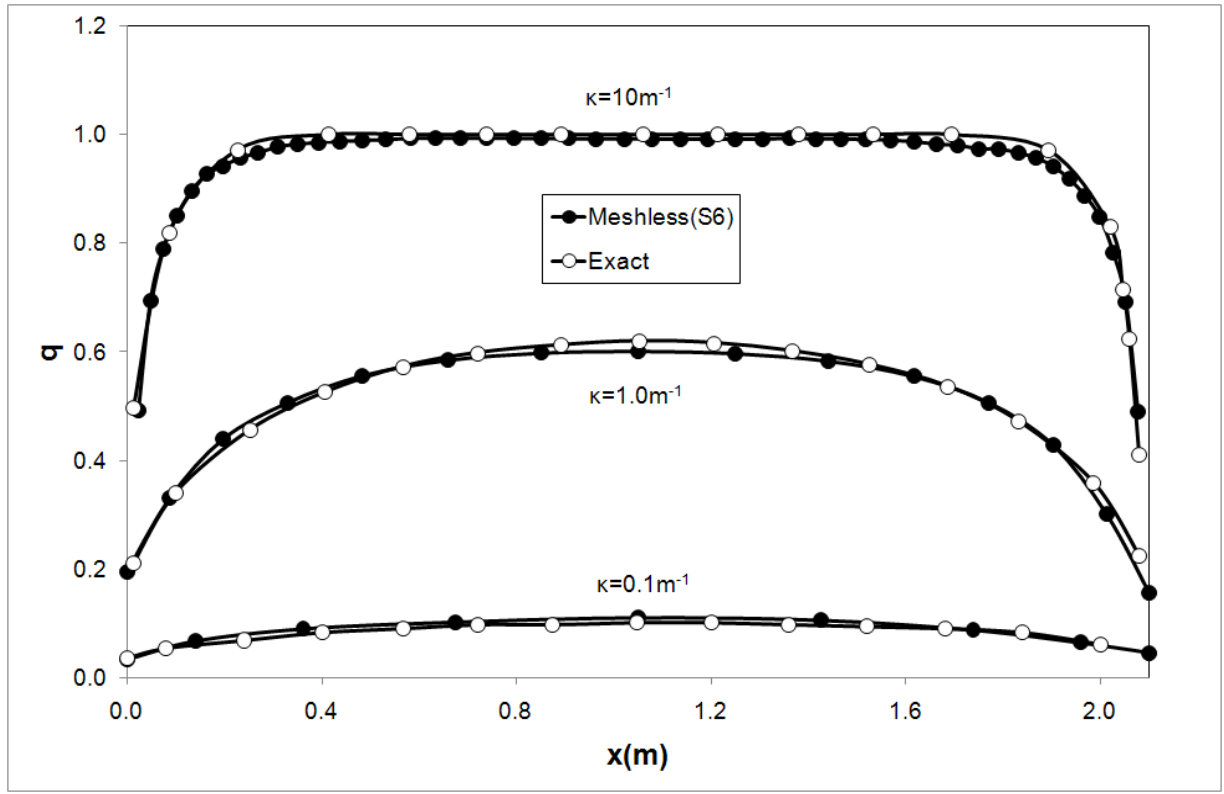

Figure 7: Comparison of radiative wall heat flux along the bottom wall

## 7.2 3D L-shaped enclosure

We consider now the 3D L-shaped enclosure containing an absorbing and emitting medium shown on Figure 8. The medium is at a temperature of 100K while the black walls are maintained at 500K. This problem has been studied by Malalasekera et al [28] by using the discrete transfer method.

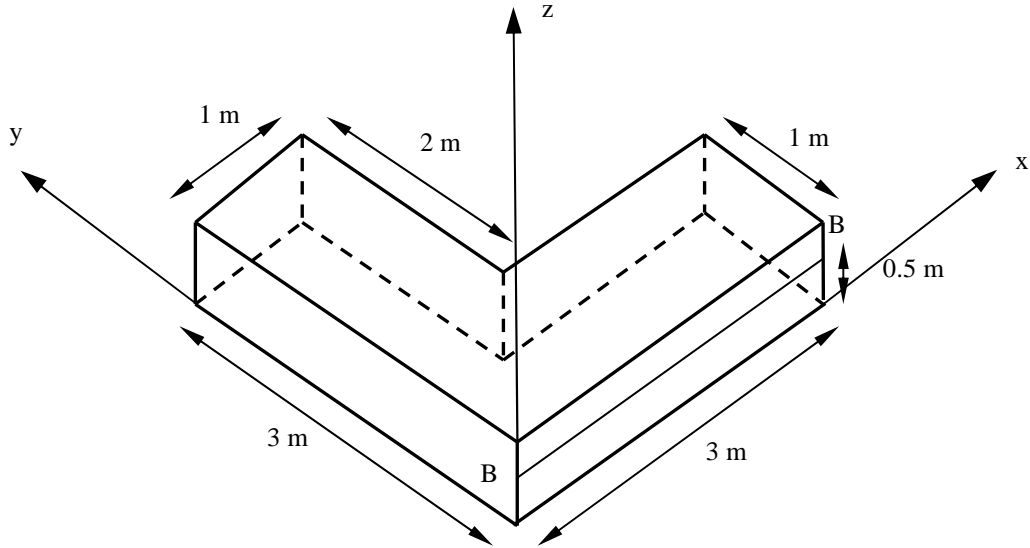

Figure 8: L Shaped geometry

We show on Figure 9 the predicted heat flux along the B-B line (marked on Figure 8) for five different absorption coefficients. Two different sets of grids are used. The first one uses 6431 nodes for $\kappa=0.5$ m$^{-1}$ and $\kappa=1$ m$^{-1}$ while the second one uses 15105 nodes for $\kappa=2$ m$^{-1}$ and $\kappa=5$ m$^{-1}$. As it can be seen, the results are in good agreement with those of Malalasekera et al [28].

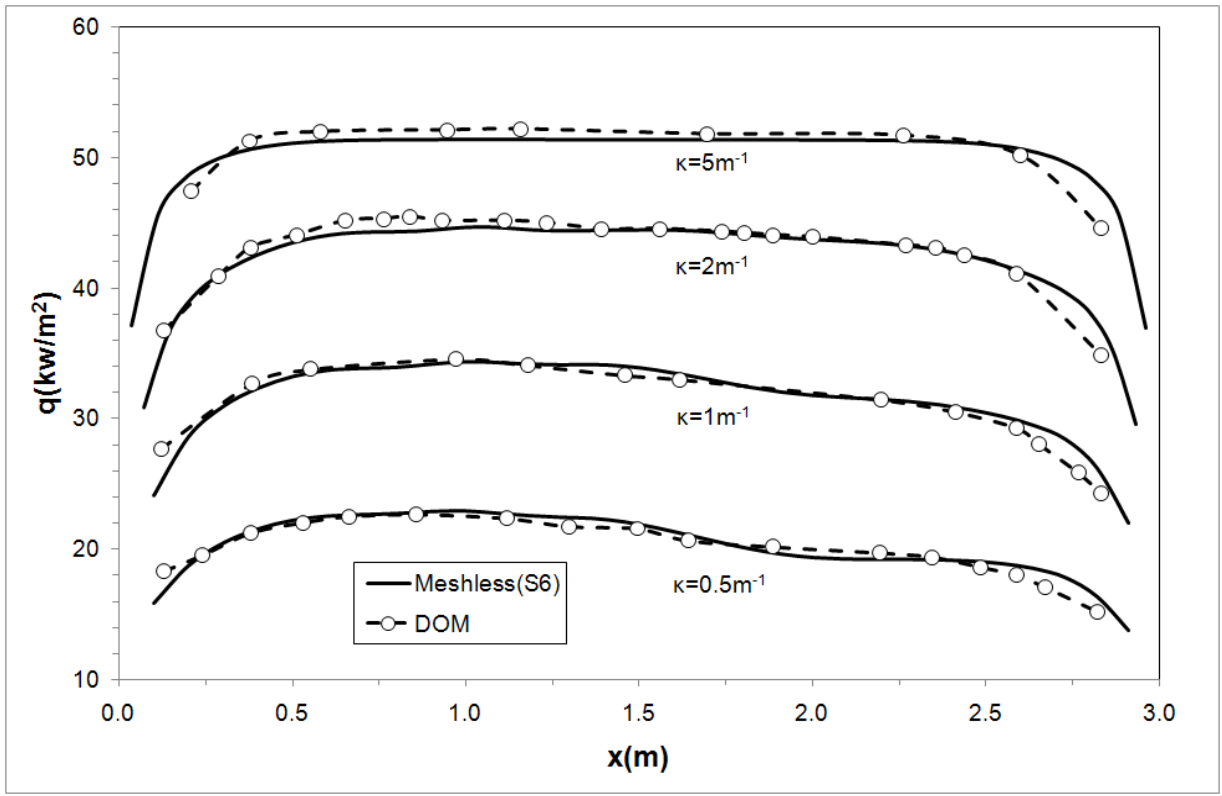

Figure 9: Net Heat flux along the line B-B of the L shaped enclosure

## 7.3 Elliptical enclosure

We consider here an absorbing-emitting medium in an elliptical enclosure with a radius R=1m, a width W=1.5m and a length H=2m respectively as shown on Figure 10. The medium is maintained at an emissive power of unity and the walls are black and cold (T=0K). Two values of the absorption coefficient are considered, namely $\kappa=1.0$ m$^{-1}$ and $\kappa=10.0$ m$^{-1}$ and an irregular grid shown on Figure 11 has been used.

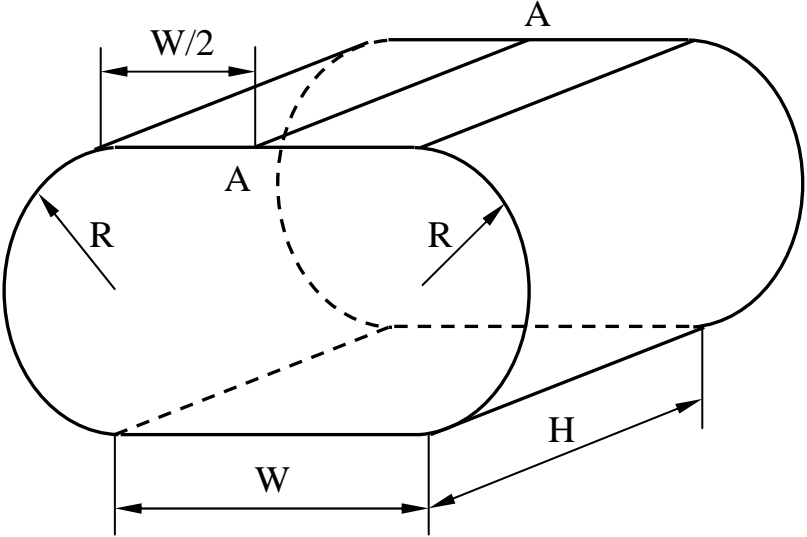

Figure 10: Elliptical enclosure

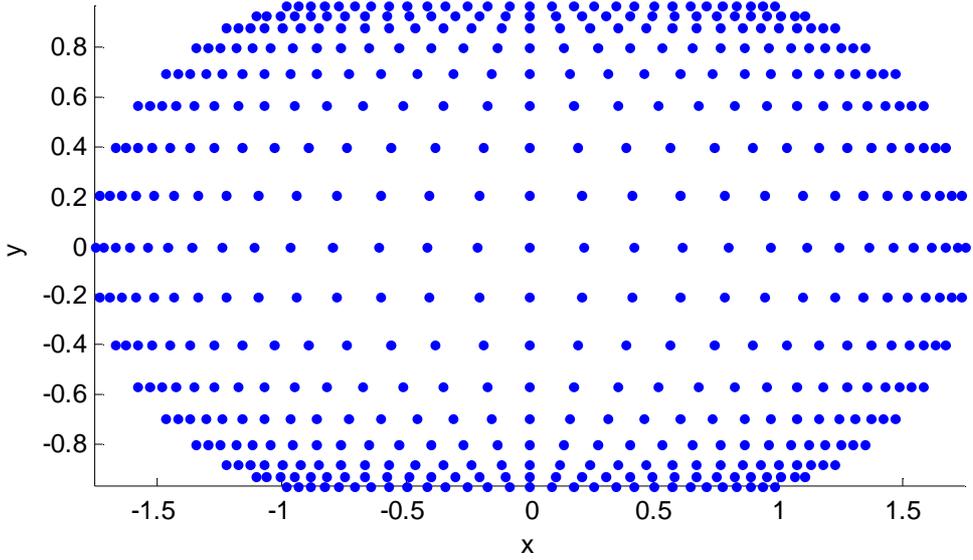

Figure 11: Example of non uniform grid

The predicted meshlesss solutions of the radiative flux along the AA line (of Figure 10) are shown on Figure 12. These solutions have been obtained with grids of 2244 nodes for $\kappa =1.0 \text{m}^{-1}$ and 6210 nodes for $\kappa =10.0 \text{m}^{-1}$ respectively. The grids were refined near the boundaries as shown on Figure 11.

As it can be seen, the meshless results are of comparable accuracy with the reference results of Chai et al.[29] and even parity formulation solved in a body-fitted system[30]

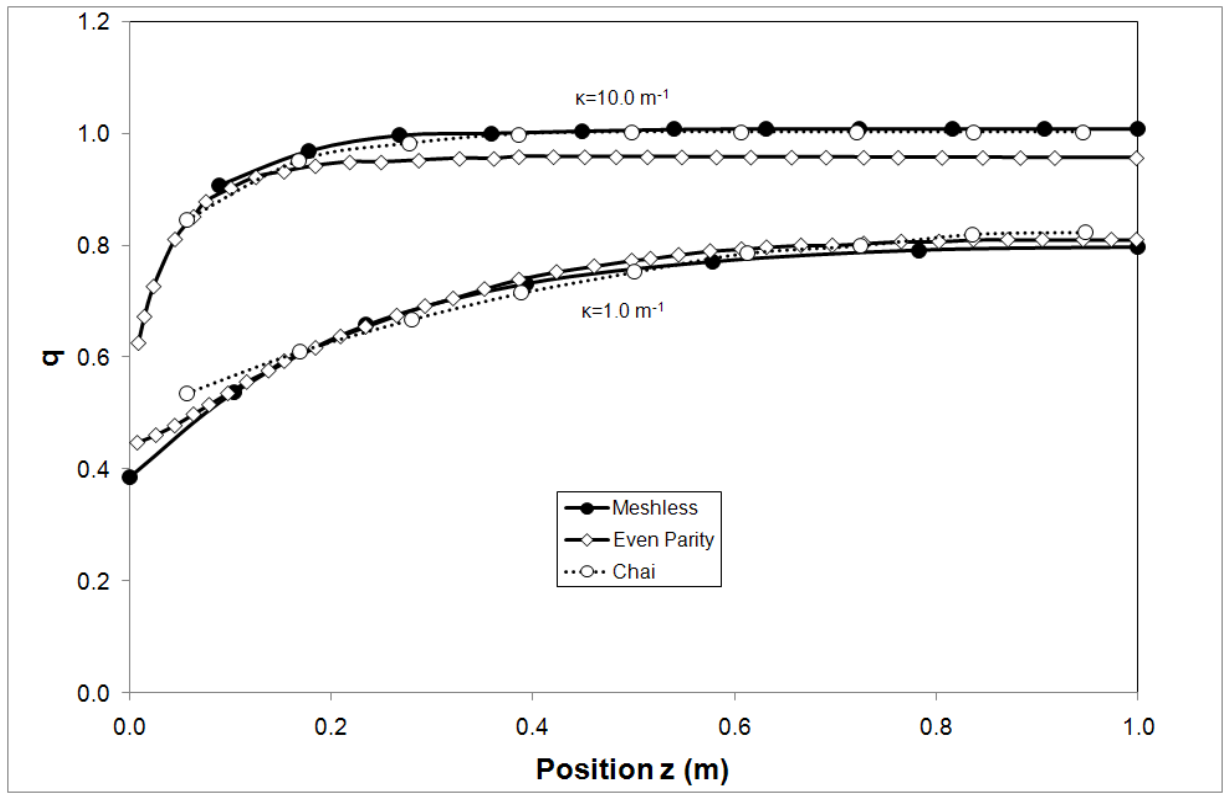

Figure 12 : Radiative flux on line AA

## 8. Conclusion

A diffuse approximation meshless method is employed for solving radiative transfer in 2D and 3D complex geometries in the even parity formulation of the discrete ordinates method. The results are compared with other benchmarks results and a good agreement has been observed. The results show that the method has a good accuracy in solving radiative heat transfer in absorbing, emitting and scattering media in complex geometry.

# References

[1] S.N.Atluri and S.Chen, The Meshless Local Petrov-Galerkin (MLPG) Method, Tech Science Press, 2002

[2] G.R. Liu, Mesh Free Methods: Moving beyond the Finite Element Method, CRC Press,Boca Raton, FL, 2002

[3] Meshfree & Particle Based Approaches in Computational Mechanics, P. Breitkopf and A.Huerta, Editors, ISTE Publishing Company, 2004

[4] P. Lancaster and K.Salkauskas, Surfaces generated by moving least squares methods, Mathematics of Computations, 37,141-158,1981

[5] B. Nayroles, G. Touzot and P. Villon, L'approximation diffuse, C. R. Acad. Sci. Paris 313, pp. 293–296, 1991

[6] B. Nayroles, G. Touzot and P. Villon, Generalizing the finite element method. Diffuse approximation and diffuse elements, Computational Mechanics,vol.10, pp.307-318, 1992

[7] T. Belytschko, Y. Y. Lu, and L. Gu, "Element free Galerkin methods, International Journal for Numerical Methods in Engineering, vol. 37, pp. 229–256, 1994

[8] Sadat H. et Prax C., Application of the diffuse approximation for solving fluid flow and heat transfer problems, International Journal Heat Mass Transfer, Vol. 39, no. 1, pp. 214-218, 1996

[9] C.Prax, P.Salagnac and H.Sadat, Diffuse approximation and control volume finite element methods, a comparative study,  Numerical Heat transfer, Part B, Fundamentals, 1996

[10] Stéphane Couturier and Hamou Sadat, Solution of Navier-Stokes equations in primitive variables by diffuse approximation, Comptes Rendus Académie des Sciences,  Vol. 326, no. 2, pp. 117-119, February 1998

[11] Couturier S. and Sadat H., A Meshless Method for Solving Incompressible Fluid Flow Problems, European Journal of Computational Mechanics, Vol.7, no. 7, 1998

[12] T.Sophy, H.Sadat and C.Prax, A meshless formulation for three dimensional laminar natural convection, Numerical Heat Transfer, Part B,Vol.41, pp. 433-445, 2002

[13] Bertrand O., Binet B., Combeau H., Couturier S., Delannoy Y., Gobin D., Lacroix M., Le Quéré P., Médale M., Mencinger J., Sadat H., Vieira G., Melting driven by natural convection : A comparison exercice, International Journal Thermal Sciences, Vol 38, pp. 5-26, 1999

[14] Raithby GD, Chui EH. A finite-volume method for predicting a radiant heat transfer in enclosures with participating media. Journal of Heat Transfer, Vol.112, pp. 415–23, 1990

[15] Chai JC, Lee HS, Patankar SV. Finite volume method for radiation heat transfer, Journal of Thermophysics and Heat Transfer, Vol.8,pp.419–25, 1994

[16] D.Lemonnier and V.LeDez, Discrete ordinates solution of radiative transfer across a slab with variable refractive index, Volume 73, Issue 2-5, Pages 195-204, March 2002

[17] An W, Ruan LM, Qi H, Liu LH. Finite element method for radiative heat transfer in absorbing and anisotropic scattering media. Journal of quantitative Spectroscopy and Radiative Transfer, Vol.96,pp.409–22, 2005

[18] Seok Hun Kang, Tae-Ho Song, Finite element formulation of the first- and second-order discrete ordinates equations for radiative heat transfer calculation in three-dimensional participating media, Journal of Quantitative Spectroscopy and Radiative Transfer, Vol.109, pp.2094–2107, 2008

[19] Cheong KB, Song TH. Examination of solution methods for the second-order discrete ordinate formulation, Numerical Heat Transfer, Part B, Vol.27,pp.155–73, 1995

[20] Fiveland W.A. and Jessee J.P., Comparaison of discrete ordinates formulations for radiative heat transfert in multidirectional geometries. Journal of Themophysics and Heat Transfer, Vol 9, n° 1, 1995.

[21] J.Y. Tan, L.H. Liu, and B.X. Li. Least-Squares Collocation Meshless Approach for Transient Radiative Transfer. Journal of Thermophysics and Heat Transfer, Vol.20, no. 4, October–December 2006

[22] L.H. Liu, J.Y. Tan.Least-squares collocation meshless approach for radiative heat transfer in absorbing and scattering media. Journal of Quantitative Spectroscopy and Radiative Transfer , Vol. 103,pp.545–557, 2007

[23] J. Y. Tan, J. M. Zhao, L. H. Liu, and Y. Y. Wang, Comparative Study on Accuracy and Solution Cost of the First/Second-Order Radiative Transfer Equations Using the Meshless Method, Numerical Heat Transfer B, vol. 55, pp. 324–337, 2009

[24] H. Sadat, On the use of a meshless method for solving radiative transfer with the discrete ordinates formulation , Journal of Quantitative Spectroscopy and Radiative Transfer, Volume 101, Issue 2, Pages 263-268, September 2006

[25] Chai, J. C. Parthasarathy, G., Lee, H. S. and Patankar, S. V., Finite volume radiative heat transfer procedure for irregular geometries, Journal of Thermophysics and Heat transfer, Vol. 9 , 1995

[26] M. Sakami, A. Charette, Application of a modified discrete ordinates method to two-dimensional enclosures of irregular geometry ,Journal of Quantitative Spectroscopy &Radiative Transfer,Vol.64,pp.275-298, 2000

[27] Baek SW, Kim MY, Kim JS. Nonorthogonal finite volume solutions of radiative heat transfer in a three dimensional enclosure, Numerical Heat Transfer (Part B), Vol.34, pp. 419–437,1998

[28] Malalasekera WMG, James EH. Radiative heat transfer calculations in three-dimensional complex geometries, Journal Heat Transfer ,Vol. 118, pp225–228, 1996

[29] Chai, J. C., Moder, J. P. and Parthasarathy, G., AIAA Paper 96-1889, 1996.

[30] J. LIU, Y. S. CHEN, Examination of conventional and even-parity formulations of discrete ordinates method in a body-fitted coordinate system, Journal of Quantitative Spectroscopy and Radiative Transfer ,Vol. 61, No. 4, pp. 417-431, 1999